\documentclass{ws-procs975x65}

\begin{document}

\title{CATCHING PHOTONS FROM EXTRA DIMENSIONS\footnote{To appear in the 
Proceedings of the
Eleventh Marcel Grossmann Meeting on General Relativity, Berlin, July 23 - 29, 2006}}

\author{A. DOBADO$^1$ and A.L. MAROTO$^2$}

\address{Departamento de  F\'{\i}sica Te\'orica,\\
 Universidad Complutense de
  Madrid, 28040 Madrid, Spain\\
$^1$E-mail: dobado@fis.ucm.es\\
$^2$E-mail: maroto@fis.ucm.es}

\author{J.A.R. CEMBRANOS}

\address{Department of Physics and Astronomy,\\
 University of California, Irvine, CA 92697, USA\\
E-mail: jruizcem@uci.edu }

\begin{abstract}
In extra-dimensional brane-world models with low tension, brane
excitations provide a natural WIMP candidate for dark matter.
Taking into account the various constraints coming from colliders,
precision observables and direct search, we explore the
possibilities for indirect search of the galactic halo branons
through their photon producing annihilations in experiments such
as EGRET, HESS or AMS2.
\end{abstract}


\bodymatter

\section{Low-tension braneworld phenomenology }\label{aba:sec1}

According to recent suggestions our universe could be a
3-dimensional brane, where the SM fields live, embedded in a
D-dimensional space-time \cite{ADD} (D = 4 + N).  The most
important parameters of this setup being the fundamental scale of
gravity in D dimensions $M_D$ (which is no longer the Planck scale
$M_P$) and the brane tension $\tau=f^4$. Besides the SM fields, 
other new excitations appear on the brane: Kaluza-Klein gravitons
\cite{KK} and brane fluctuations $\pi^\alpha,\alpha=1,2,... N$,
where $N$ is smaller or equal than the number of extra dimensions
\cite{BR}. These branons are the Goldstone Bosons associated to
the spontaneous breaking of the translational invariance in the
extra dimensions induced by the presence of the brane. However, in
the general case, translational invariance is not an exact
symmetry of the bulk space, i.e: branons acquire a mass M. For $f
\ll M_D$ (low tension), KK gravitons decouple from the SM
particles. Consequently, at low energies the only relevant degrees
of freedom are the SM particles and the branons whose interactions
can be described by the effective Lagrangian:
\begin{eqnarray}
{\cal L}_{Br}&=& \frac{1}{2}g^{\mu\nu}\partial_{\mu} \pi^\alpha
\partial_{\nu} \pi^\alpha-\frac{1}{2}M^2 \pi^\alpha \pi^\alpha
+ \frac{1}{8f^4}(4\partial_{\mu} \pi^\alpha
\partial_{\nu} \pi^\alpha-M_{\alpha\beta}^2 \pi^\alpha \pi^\beta g_{\mu\nu})
T^{\mu\nu}_{SM}\nonumber \\\label{lag}
\end{eqnarray}
in which, one can see that branons interact by pairs with the SM
and with a coupling controlled by the brane tension scale $f$.  For
simplicity, we assume that all branons have the same mass,
$M_{\alpha\beta}\equiv M\delta_{\alpha\beta}$. Therefore branons
are a kind of new scalar fields, whose properties (stability, weak
couplings and masses) coincide with those expected for a WIMP
(Weakly Interacting Massive Particle) \cite{CDM}.

From the above effective Lagrangian it is possible to obtain the
branon production cross sections for different colliders \cite{ACDM},  the
typical signature being missing energy and missing $P_T$, and thus to
find bounds on the $f$ and $M$ parameters for different values of
$N$. Other constraints can also be obtained by computing the
effect of virtual branons on various precision observables
\cite{CDM2} including the muon $g-2$ measurements. Taking all this into 
account, one can calculate the rate for direct detection of branons
in the current and future experiments designed for WIMP detection.
Remarkably these particles can be well accommodated within all
these bounds and still they offer definite predictions for future
direct search experiments \cite{CDM}. In addition WIMPs are expected to
annihilate in the galactic halo producing photons in different
ways. Such photons could be caught by detectors on Earth or in space, thus
providing a new indirect way to detect their presence which could
nicely complement the above mentioned more direct searches. In the
following we analyze the potential detection of these photons
coming from the galactic halo branons.

\section{Gamma rays from branon annihilation}

The photon flux in the direction of the galactic center coming
from dark matter annihilations can be written as \cite{BUB,FMW}:
\begin{eqnarray}
\frac{d\,\Phi_{\gamma}^{DM}}{d\,\Omega\,d\,E_{\gamma}} =
\frac{J_0}{N M^2}\sum_i\langle\sigma_i v\rangle
\frac{d\,N_\gamma^i}{d\,E_{\gamma}}\label{flux}
\end{eqnarray}
where $J_0$ is the integral of the dark matter mass density
profile, $\rho(r)$, along the path between the galactic center and
the gamma ray detector:
\begin{eqnarray}
J_0= \frac{1}{4\pi}\int_{path} \rho^2\,d\,l ,
\end{eqnarray}
$N$ is the number of dark matter species with mass $M$ and
$\langle\sigma_i v\rangle$ is the  thermal average of the
annihilation cross section of two dark matter particles into
another two particles. On the other hand,  the continuum photon
spectrum from the subsequent decay of particles species $i$ presents a
simple description in terms of the photon energy normalized to the
dark matter mass, $x=E_{\gamma}/M$. Thus, for each channel $i$, we have:
\begin{eqnarray}
\frac{d\,N_\gamma^i}{d\,x}= M\frac{d\,N_\gamma^i}{d\,E_{\gamma}}=
\frac{a^i}{x^{3/2}}e^{-b^ix}.
\end{eqnarray}
where $a_i$ and $b_i$ are constants. 
In the case of heavy branons, if we neglect   three
body decays and  direct production of two photons, the 
main contribution to the photon flux comes from branon annihilation into 
$ZZ$ and $W^+ W^-$. The contribution from heavy fermions, i.e. annihilation
in top-antitop can be shown to be subdominant.
The concrete
values for the above constants in those channels are: 
$a^{ZZ}=a^{W^\pm W^\mp}=0.73$
and $b^{ZZ}=b^{W^\pm W^\mp}=7.8$ \cite{BUB,FMW}. 

On the other hand, the thermal averaged cross-section $\langle\sigma_{Z,W} v\rangle$
which enters in eq. (\ref{flux}) has been
calculated in \cite{CDM} and in the non-relativistic limit is given by:
\begin{eqnarray}
\langle\sigma_{Z,W} v\rangle=\frac{M^2\sqrt{1-\frac{m_{Z,W}^2}{M^2}}
(4M^4-4M^2m_{Z,W}^2+3m_{Z,W}^4)}{64f^8\pi^2}
\end{eqnarray}

The produced high-energy gamma photons could be in the range
(30 GeV-10 TeV), detectable 
by Atmospheric
Cerenkov Telescopes (ACTs)  such as HESS, VERITAS or MAGIC. 
On the contrary, if $M<m_{Z,W}$, the
annihilation into W or Z bosons is kinematically forbidden and it is
necessary to take into account the rest of  channels, mainly 
annihilation into the heaviest possible quarks \cite{Silk}. 
In this case, the 
photon fluxes would be in the range detectable by 
space-based gamma ray observatories \cite{progress} 
such as EGRET, GLAST or AMS, 
with better sensitivities around 30 MeV-300 GeV.

{\em Acknowledgments:} This work has been partially supported by
DGICYT (Spain) under project numbers FPA 2004-02602 and FPA
2005-02327, by  NSF CAREER grant
No.~PHY--0239817, NASA Grant No.~NNG05GG44G, the Alfred P.~Sloan
Foundation and Fulbright-MEC award.



\end{document}